\documentstyle[12pt]{article}

\begin{document}

\title{Semiquantum Chaos in the Double-Well}

\vskip 0.3cm
\author{
Thomas C. Blum
\thanks{E-mail: tblum@wind.phy.bnl.gov}\\
Physics Department, 510A\\
Brookhaven National Laboratory \\
Upton, NY 11973-5000, USA
\and
Hans-Thomas Elze
\thanks{E-mail: elze@cernvm.cern.ch,
hans-thomas.elze@physik.uni-regensburg.de}\\
Institut f\"ur Theoretische Physik, Universit\"at Regensburg\\
93053 Regensburg, Germany\\
Physics Department, University of Arizona\\
Tucson, AZ 85721, USA
}

\maketitle

\begin{abstract}
\renewcommand{\baselinestretch}{1.2}

The new phenomenon of semiquantum chaos is analyzed in a classically
regular double-well oscillator model. Here it arises from a doubling
of the number of effectively classical degrees of freedom, which are
nonlinearly coupled in a Gaussian variational approximation (TDHF) to
full quantum mechanics. The resulting first-order nondissipative
autonomous flow system shows energy dependent transitions between
regular behavior and semiquantum chaos, which we monitor by Poincar\'e
sections and a suitable frequency correlation function related to the
density matrix. We discuss the general importance of this new form of
deterministic chaos and point out the necessity to study open
(dissipative) quantum systems, in order to observe it experimentally.
\end{abstract}
\vskip 0.3cm
\begin{center}
\end{center}
\vskip 1.5cm
\hfill\parbox{5cm}{TPR-95-28\\nlin-sys/9511007\\
}

\newpage

\section{Introduction}

Recently the concept of \underline{semiquantum chaos} has been
introduced in order to characterize a particular form of deterministic
chaos \cite{Schuster}. Namely, dynamical systems with both quantum and
classical degrees of freedom may show irregular behavior due to their
generic coupled nonlinearities \cite{Cooper,Pattanayak,Jona-Lasinio}.
The appearance of both quantum and classical degrees of freedom (d.o.f.)
generally can be thought of as a more or less valid approximation,
depending on the physical circumstances, describing a truly complex
nondissipative quantum system.

For example, the authors of Ref. \cite{Cooper} considered the zero
momentum (long wavelength) part of the problem of pair production of
charged scalar particles by a strong external electric field.
In this limit the problem could be reduced to a classical oscillator
interacting with a quantum mechanical one through a biquadratic
coupling; i.e. a two d.o.f. system, one classical and the other purely
quantum. They represent the electromagnetic vector potential and the
charged matter field, respectively. In this case and for a physically
motivated subset of all possible initial conditions, the dynamics can
be mapped onto an equivalent completely classical problem with two
effective d.o.f. Their motion is described by two coupled nonlinear
second-order differential equations, which yield regular or irregular
trajectories depending on initial conditions and suitably chosen model
parameters. Clearly, an approximation to a purely quantum system, a
quantum field theory in fact, has been treated here and eventually
gives rise to \underline{deterministic chaos in an equivalent
low-dimensional classical system.}

The latter observation presents the starting point of our investigation
of semiquantum chaos in the one-dimensional double-well problem. It can
be considered as the zero-dimensional limit of the scalar Higgs field
model \cite{Ryder}. To study the entropy production  and thermalization
in strong interactions at ultrarelativistic energies a related scalar
quantum field theory has recently been treated in the time-dependent
Hartree-Fock approximation (TDHF) equivalent to a Gaussian variational
principle \cite{Elze,Jackiw} (see also further references therein).
The resulting field equations in the long wavelength limit, i.e.
neglecting spatial variations, reduce to a coupled pair of Hartree-Fock
type equations for one-dimensional double-well oscillators. They represent
a semiclassical approximation to the full quantum problem in terms of
spatially homogenous fast and slow modes, respectively. The equations
for a single such oscillator will be given in Section II and their
solutions studied in detail in the following.

At this point it seems sufficient to state that mean-field type TDHF
quantum corrections to the classical equation for the double-well
oscillator, which behaves regularly in the classical limit, introduce
characteristic nonlinear features including additional effectively
classical d.o.f. They, however, describe quantum fluctuations of the
basic coordinates and conjugate momenta. Such additional d.o.f. are
an essential element to allow for the possibility of semiquantum chaos.
We recall that a classical one-dimensional anharmonic oscillator, which
can be described by a set of two autonomous first-order differential
equations, cannot have chaotic trajectories according to the
Poincar\'e-Bendixson theorem \cite{Schuster}, which rules out chaotic
flow in a bounded region in two-dimensional (phase) space.

The time-dependent variational principle \cite{Jackiw} using generalized
Gaussian trial wave functions has, in fact, already been applied in
quantum chemistry for quite a while (see Refs. \cite{Heller,Heller2}
for some early and more recent work, respectively, citing numerous
further relevant references). However, it has only recently been pointed
out that the parameters (parameter functions in field theory \cite{Elze})
specifying the Gaussian wave functions (functionals in field theory)
are governed by a set of equations according to the variational principle
which describe an equivalent classical Hamiltonian system
\cite{Pattanayak}. It is this equivalent classical system which may show
what has been termed semiquantum chaos, even if the subset of equations
belonging to the proper classical limit (neglecting quantum fluctuation
d.o.f.) produces only strictly regular motion. Furthermore, it is
generally believed that the full quantum dynamics obeying the
Schr\"odinger wave equation, which is a linear partial differential
equation (however, a functional l.p.d. equation in field theory),
is always regular (see Refs. \cite{Pattanayak,Heller2} for further
discussions on this point). Thus, the question of \underline{quantum chaos}
usually pertains to more subtle aspects of the dynamics which reflect
some traces of deterministic chaos in the corresponding classical system
\cite{Schuster,Heller2}, if at all.

Presently, we want to address the question instead, rather independently
of the behavior of the full quantum dynamics or even a regular classical
limit thereof, whether or how the onset of semiquantum chaos is related
to the breakdown of the underlying semiclassical approximation.
Correspondingly, one may reconsider the often posed problem of whether or
where in the relevant classical phase space there are regions in which
the system can safely be or even has to be studied using a semiclassical
approach \cite{Carruthers}.

Finally, we want to draw attention to Ref. \cite{Jona-Lasinio}. There,
the possibility of chaotic behavior without classical counterpart is
pointed out in a many-body system undergoing multiple resonant tunneling.
More specifically, an electron cloud in a three-well heterostructure
was represented by an effective one-particle wave function obeying a
nonlinear Hartree equation. Clearly, the inherently semiclassical
description gives rise to the essential nonlinearity of the second-order
partial differential equation again. However, the authors of this
work speculate that the many-body character of the system, i.e. infinitely
many d.o.f. already on the classical field level, might stabilize the
reported semiquantum chaos in this case to persist even in a fully quantum
treatment. This necessitates a time-dependent approach to quantum field
theory, which has not been worked out for the nonrelativistic problem
at hand to date. We will come back to this in the final section where we
discuss one possible first step beyond our present simple double-well
oscillator model in the direction of a quantum field theory,
which is motivated by the above mentioned issues
\cite{Jona-Lasinio,Elze}. In any case, it is gratifying that at least
for solid state heterostructures an experimental verification of
semiquantum chaos seems within reach (see further references in
\cite{Jona-Lasinio}.

Our paper is organized as follows. Section II defines the model and
describes briefly the time-dependent variational principle, from which
we obtain our basic set of equations. In Section III we present some
analytical considerations and introduce the effective semiquantum
potential, which should help to visualize the results of our numerical
studies presented in Section IV. In Section V we briefly compare the
semiquantum dynamics with exact solutions of the Schr\"odinger equation
and summarize our results. We point out some interesting problems
related to environment-induced decoherence and chaotic entropy
production in open quantum systems, which are left for future work.

We discuss qualitatively there, how semiquantum chaos can be
reconciled with a regular classical limit and the intrinsic linearity
of quantum mechanics, in particular.

\newpage

\section{The Semiclassical Double-Well Oscillator}

The purpose of this section is to introduce our toy model of the
simple one-dimensional double-well oscillator which, however, will be
studied in a nonperturbative semiclassical approximation (TDHF, cf.
below). As explained in the Introduction, we consider this as a
necessary first step towards the study of a (3+1)-dimensional
field theory \cite{Elze}. The classical action of our model is given by
\begin{equation}
\label{g1}
S=\int dt\;\lbrace\frac{1}{2}(\partial _{t}\varphi)^{2}-v(\varphi)
\rbrace\;\; ,
\end{equation}
\noindent
with
\begin{equation}
\label{g2}
v(\varphi)\equiv -\frac{1}{2}\mu^{2}\varphi^{2}+\frac{\lambda}{4!}
\varphi^{4}\;\; ,
\end{equation}
\noindent
and $\varphi\equiv\varphi(t)$ denotes the oscillator coordinate, which is
the analogue of a scalar field, $\varphi\equiv\varphi (x,t)$, in
zero space dimensions. Note that the potential includes a negative
``mass'' term $\sim\mu^{2}$, which allows for spontaneous symmetry
breaking in the anharmonic potential $\sim \varphi^4$. Fig. 1 shows
the potential $v(\varphi)$ for the typical model parameters used
throughout this work, $\mu^{2}=1.0, \lambda =0.06$ (we employ units such
that $\hbar=c=1$) together with a typical classical phase space
trajectory oscillating in one of the wells.

The Schr\"odinger equation following from eq. (\ref{g1}) (coordinate
representation),
\begin{equation}
\label{g3}
i\partial _{t}\psi (\varphi;t) = \hat{H}\psi (\varphi;t) = \lbrace -
\frac{1}{2}\frac
{d^2}{d\varphi^2}+v(\varphi)\rbrace\psi (\varphi;t)\;\; ,
\end{equation}
\noindent
is linear in the wave function $\psi$. Thus, there is no mixing of its
Fourier components,
\begin{equation}
\label{g4}
\psi (\varphi;\omega )\equiv\int\limits^{\infty}_{-\infty}dt\;
\mbox{e}^{-i\omega t}\psi (\varphi;t)\;\; ,
\end{equation}
\noindent
which allows a trivial decomposition of the full quantum dynamics
w.r.t. fast and slow modes. This explains why at first sight one does
not expect quantum chaos, no matter whether the corresponding classical
system behaves regularly or not, and why many efforts have been made to
find traces of classical chaos in the fully quantized system (see, e.g.,
Refs. \cite{Schuster,Heller2,Ott} and further references therein).

We remark here that the situation for the field theory functional
Schr\"odinger equation, despite its apparent similarity to eq.
(\ref{g3}), seems even more complex, since the relation of the physical
(most commonly employed single-particle) observables to the wave
functional is more complicated in this case. This is exemplified already
by the quantum Brownian motion of a single particle interacting with a
quantized electromagnetic field, phonon, or other radiation field
\cite{Elze,Elze2,Grabert}, cf. Section V.

Presently, we do not follow the established routes of investigations
into quantum chaos. Rather, we study chaotic behavior generated in the
semiclassical regime even for classically regular systems, such as
our model defined by eqs. (\ref{g1}), (\ref{g2}). Before we derive the
relevant equations of motion here, we pause to consider the observables
of our model in more detail. Apart from the usual expectation values of
the coordinate and its conjugate momentum,
\begin{equation}
\label{g5}
\bar{\varphi} (t)\equiv\langle\hat{\varphi }\rangle\;\; ,\;\;\;
\bar{\pi} (t)\equiv\langle\hat{\pi}\rangle =
\langle -i\frac{d}{d\varphi }\rangle\;\; ,
\end{equation}
\noindent
we will be particularly interested in expectation values of functions
of $\varphi$,
\begin{equation}
\label{g6}
O_\varphi (t)\equiv\langle O(\hat{\varphi})\rangle =
\int\limits_{-\infty}^{\infty}
d\varphi\;\psi (\varphi;t)O(\varphi)\psi^\ast (\varphi;t)\;\; ,
\end{equation}
\noindent
i.e. fluctuation variables. By Fourier transformation we obtain
\begin{equation}
\label{g7}
O_{\varphi}(\omega) = \int\limits_{-\infty}^{\infty}d\varphi
\int\limits_{-\infty}^{\infty}\frac{d\omega'}{2\pi}\;\psi (\varphi;\omega')
O(\varphi)\psi^\ast (\varphi;\omega'-\omega)\;\; .
\end{equation}
\noindent
Thus, we have to find the spectrum of frequencies in
\begin{equation}
\label{g8}
\rho (\varphi,\varphi;\omega)\equiv\int\limits_{-\infty}^{\infty}
\frac{d\omega'}{2 \pi}\;\psi (\varphi;\omega') \psi^\ast
(\varphi;\omega'-\omega)\;\; ,
\end{equation}
\noindent
which is the Fourier transform of a diagonal element of the density
matrix which pertains to the pure state $\vert\psi (t)\rangle$
(coordinate representation). For simplicity we do not consider the
off-diagonal density matrix elements at present, which would be needed to
evaluate more general observables
$\langle O(\hat{\varphi},\hat{\pi})\rangle$. Clearly, the diagonal elements
$\rho (\varphi,\varphi;\omega)$
yield a frequency correlation function defined at each point
$\varphi$, which presents a simple and particularly useful example out
of a large variety of higher order correlation functions.

In order to illustrate the information provided by the frequency
correlation function, we evaluate it formally in the full quantum
case. Using the fact that the spectrum of the double-well
Schr\"odinger equation (\ref{g3}) corresponds to discrete stationary
bound states,
\begin{equation}
\label{g9}
\psi_{n} (\varphi;t) = \phi_{n} (\varphi)\mbox{e}^{i \omega_{n} t}\;\; ,
\end{equation}
\noindent
gives
\begin{equation}
\label{g10}
\psi^{(\ast )}_n(\varphi;\omega ) = 2\pi\phi_{n}^{(\ast )}(\varphi)\delta
(\omega-\omega_{n})\;\; .
\end{equation}
\noindent
A general pure state is
\begin{equation}
\label{g11}
\psi (\varphi; \omega) = 2\pi\sum\limits_{n}c_n\phi_n (\varphi)
\delta (\omega-\omega_n)\;\; .
\end{equation}
\noindent
Then, we obtain from eqs. (\ref{g8}) and (\ref{g11})
\begin{equation}
\label{g12}
\rho (\varphi,\varphi;\omega) = \sum\limits_{n,n'} M_{nn'}
(\varphi)\delta (\omega-\omega_n + \omega_n')\;\; ,
\end{equation}
\noindent
i.e. a spectrum of discrete lines arising at all
$\omega = \omega_n-\omega_n'$, with a strength
\begin{equation}
\label{g13}
M_{nn'} (\varphi)\equiv 2\pi c_n c_{n'}^\ast\phi_n (\varphi)\phi_{n'}^\ast
(\varphi) = M_{n'n}^\ast\;\; .
\end{equation}
\noindent
The strength matrix $M$ is Hermitean and time-independent, implying
constant real eigenvalues. In particular,
\begin{eqnarray}
\label{g14}
\rho (\varphi,\varphi;\omega = 0) &=& \mbox{const}\cdot\mbox{Tr} M(\varphi)
= \int\limits_{-\infty}^{-\infty}dt\;\psi (\varphi;t)\psi^\ast
(\varphi;t) \nonumber \\
				  &=& \mbox{const}\cdot 2\pi
\sum\limits_{n}
\vert c_n\vert ^2\vert\phi_n (\varphi)\vert ^2\;\; .
\end{eqnarray}
\noindent
Thus, the strengths of the lines depend on the actual state under
consideration, i.e. the amplitudes $c_n$, and are completely
determined by the initial condition, $\psi (\varphi;t=0)$, whereas
their positions are given by the distribution of the energy level
spacings of the system. We will employ the frequency correlation
function as a diagnostic tool in Section IV to monitor the onset of
semiquantum chaos, in which case the frequency correlation function
becomes increasingly noisy with its discrete line character eventually
disappearing completely.

Next, we turn to the derivation of appropriate semiquantum equations
of motion. The TDHF or Gaussian variational
approximation can be easily obtained with the help of Dirac's
variational principle \cite{Jackiw}. The latter can be stated as follows:
\begin{equation}
\label{g15}
\frac{\delta\Gamma[\psi ]}{\delta\psi} = 0\;\; ,\;\;\;\mbox{for all}\;\psi
\;\;\mbox{with}\;\langle\psi\vert\psi\rangle = 1\;\; ,
\end{equation}
\noindent
and
\begin{equation}
\label{g16}
\Gamma[\psi ] = \int dt\;\langle\psi (t)\vert (i\partial_t - \hat{H})\vert
\psi (t)\rangle\;\; ,
\end{equation}
\noindent
i.e. requiring the effective action $\Gamma$ defined in eq. (\ref{g16})
to be stationary against arbitrary variations of the normalized wave
function, which vanish at $t \rightarrow \pm \infty$, is equivalent to
the exact Schr\"odinger equation (3). With the variational principle
one can solve the quantum mechanical time-evolution problem approximately
by restricting the variation of the wave function to a subspace of the
full Hilbert space.

In the following we work with properly normalized Gaussian
trial wave functions,
\begin{eqnarray}
\psi_{G} (\varphi;t) &\equiv& (2\pi G(t))^{-\frac{1}{4}}\exp\{
-[\frac{1}{4}G^{-1}(t)-i\sigma (t)][\varphi-\bar{\varphi}(t)]^2
\nonumber \\
		     &\;&\;\;\;\;\;\;\;\;\;\;\;\;\;\;\;\;\;\;\;
		     +i\bar{\pi}(t)[\varphi-
\bar{\varphi}(t)]\}\;\; ,
\label{g17}
\end{eqnarray}
\noindent
which leads to the TDHF approximation via the evaluation of the
effective action, eq. (\ref{g16}), i.e. by performing simple Gaussian
integrations. Thus, we obtain:
\begin{eqnarray}
\Gamma[\psi_{G}] &=& \int dt\left\{\bar{\pi}\dot{\bar{\varphi}}-
\frac{1}{2}\bar{\pi}^2 - v(\bar{\varphi})\right. \label{g18} \\
		 &\;&\left. +\hbar [\sigma\dot{G}-2\sigma^2G-\frac{1}{8}
G^{-1}-\frac{1}{2!} v''(\bar{\varphi})G]-\frac{3}{4!}\hbar^2v''''
(\bar{\varphi})G^2\right\}\;\; , \nonumber
\end{eqnarray}
\noindent
with $v''(\bar{\varphi})=-\mu^2+\frac{1}{2}\lambda\bar{\varphi}^2$,
$v''''(\bar{\varphi}) = \lambda$ from eq. (\ref{g2}).
In eq. (\ref{g18}) we inserted the appropriate factors of $\hbar$
which exhibit the classical and semiquantum contributions to the
effective action. Clearly, the dynamics is now described by the four
time-dependent functions which parametrize the trial wave function,
eq. (\ref{g17}). Here $\bar{\varphi}$ and $G$ play the role of the
effective coordinates, while $\bar{\pi}$ and $\sigma$ present the
conjugate momenta. This is further illustrated by the expectation values,
\begin{equation}
\langle\varphi\rangle_G=\bar{\varphi}(t)\;\; ,\;\;\;
\langle -i\frac{d}{d\varphi}\rangle_G=\bar{\pi}(t)\;\;,
\label{g19}
\end{equation}
\noindent
and
\begin{equation}
\langle\varphi^2\rangle_G=\bar{\varphi}^2(t)+G(t)\;\; ,\;\;\;
\langle i\partial_{t}\rangle =\bar{\pi}(t)\dot{\bar{\varphi}}(t)-
\dot{\sigma}(t)G(t)\;\; ,
\label{g20}
\end{equation}
\noindent
which are calculated using eq. (\ref{g17}). Furthermore,
\begin{eqnarray}
(\langle\varphi^2\rangle_G-\bar{\varphi}^2)^{\frac{1}{2}}\cdot
(\langle\pi^2\rangle_G-\bar{\pi}^2)^{\frac{1}{2}}&=&
	G^{\frac{1}{2}}\cdot (\frac{1}{4}G^{-1}+4\sigma^2G)^{\frac{1}{2}}
\nonumber \\
    &=&\frac{1}{2}(1+(4\sigma G)^2)^{\frac{1}{2}}\geq\frac{1}{2}\;\; ,
\label{g20.1}
\end{eqnarray}
\noindent
which demonstrates the uncertainty relation in the present context.
Minimum uncertainty coherent states yield the lower bound on the
r.h.s. of eq. (\ref{g20.1}).

Finally, the \underline{semiquantum equations of motion} are obtained as
the Euler - Lagrange equations for the effective action, eq. (\ref{g18}),
by independent variations w.r.t. $\bar{\varphi}$, $\bar{\pi}$, $G$ and
$\sigma$:
\begin{eqnarray}
\dot{\bar{\varphi}}&=&\bar{\pi}\;\; ,
\label{g21} \\
\dot{\bar{\pi}}&=&-v'(\bar{\varphi})-\frac{\hbar}{2}v'''(\bar{\varphi})G
\;\; ,
\label{g22} \\
\dot{G}&=&4\sigma G\;\; ,
\label{g23} \\
\dot{\sigma}&=&-2\sigma^2+\frac{1}{8}G^{-2}-\frac{1}{2}v''
(\bar{\varphi})-\frac{\hbar}{4}v''''(\bar{\varphi})G\;\; ,
\label{g24}
\end{eqnarray}
\noindent
with $v'(\bar{\varphi})=-\mu^2\bar{\varphi}+\frac{\lambda}{3!}$,
$v'''(\bar{\varphi})=\lambda\bar{\varphi}$, and $v''$, $v''''$
as given after eq. (\ref{g18}). Several remarks are in order here:
\begin{itemize}
\item[$\bullet$] Equations (\ref{g21})-(\ref{g24}) present a coupled set
of four autonomous first-order nonlinear differential equations, which
explains the potential for semiquantum chaos of the double-well
oscillator, for example.
\item[$\bullet$] Equivalently, by eliminating the momenta $\bar{\pi}$
and $\sigma$, one obtains two coupled second-order equations, which bear
some characteristic resemblance to eqs. (16) in Ref. \cite{Cooper},
i.e. the particle production problem mentioned in the Introduction.
\item[$\bullet$] It follows from eq. (\ref{g18}) that only eqs. (\ref{g21}),
(\ref{g22}) survive the simple-minded classical limit with $\hbar
\rightarrow 0$; thus, eqs. (\ref{g23}), (\ref{g24}) present the dynamics
of an additional effectively classical d. o. f., which arises through
the semiclassical approximation of (Gaussian) quantum fluctuations.
\end{itemize}

The TDHF approximation in the form of eqs. (\ref{g21}) - (\ref{g24})
was previously studied for various quantum mechanical models in Ref.
\cite{Cooper2}; the potential for semiquantum chaos, however, has
only recently been noticed based on a derivation via Ehrenfest's theorem
\cite{Pattanayak}. Note that the above equations of motion decouple and
can be solved analytically, of course, in the harmonic limit, $v''=
\mbox{const}$, $v'''=v''''=0$.

To conclude this section, note that the frequency correlation
function,
eq. (\ref{g8}), in TDHF approximation is given by
\begin{equation}
\rho (\varphi,\varphi;\omega) = \int\limits_{-\infty}^{\infty}
dt\;\mbox{e}^{-i\omega t}(2\pi G)^{\frac{1}{2}}\exp\{-\frac{1}{2} G^{-1}
[\varphi - \bar{\varphi}]^2\}\;\; ,
\label{g25}
\end{equation}
\noindent
using eq. (\ref{g17}). Clearly, the magnitude of $G$ determines the
admissible amount of quantum fluctuation around the classical
expectation value $\bar{\varphi}$, cf. (\ref{g19}), such that the integrand
in eq. (\ref{g25}) is not negligible. Thus, the relevant contributions
to $\rho (\varphi,\varphi;\omega)$ come from those parts of a
semiquantum trajectory of the system, i.e. $\{\bar{\varphi(t)},G(t)\}$,
which are sufficiently close to the point $\varphi$ and have a time
dependence characterized approximately by the frequency $\omega$.

\section{Energy, Effective Potential and Instabilities}

The purpose of this section is to report some analytical considerations,
which may help to better understand the dynamics of the semiquantum
equations of motion, eqs. (\ref{g21}) - (\ref{g24}).

To begin with, the Hamiltonian contribution to the effective action,
cf. eqs. (\ref{g16}), (\ref{g18}) together with the second of eqs.
(\ref{g20}), yields the conserved total energy for our nondissipative
model. The resulting expression is still a rather complicated function
of $\bar{\varphi}$, $\bar{\pi}$, $G$, and $\sigma$. For simplicity, we
choose \underline{initial conditions} with zero momenta in the
following, $\bar{\pi}(t=0)=\sigma (t=0)=0$. Then, evaluating the total
energy, E$_0$, at the minimum of the classical potential, for example,
we obtain
\begin{equation}
E_0(G)=\frac{1}{8}G^{-1}-\frac{3}{2}\frac{\mu^4}{\lambda}+
\mu^2G+\frac{1}{8}\lambda G^2\;\; .
\label{g26}
\end{equation}
\noindent
With all other parameters kept fixed, the initial energy or,
equivalently, the initial value $G(t=0)$ is the relevant \underline{control
parameter} for our model.

Furthermore, comparing the initial value $E_0(G)$ with the total
energy evaluated at any later time, we obtain a simple measure for the
numerical accuracy of the simulations to be discussed in Section IV.

Secondly, in analogy to a classical system in two dimensions
(``coordinates'' $\bar{\varphi}$ and $G$), we consider the
\underline{effective potential} which is defined as the total minus
the kinetic energy, i.e. $V(\bar{\varphi}, G)\equiv
E(\bar{\varphi},G;\bar{\pi}=\sigma =0)$, in terms of the total
energy. Thus,
\begin{equation}
V(\bar{\varphi},G)=v(\bar{\varphi})+\hbar [\frac{1}{8}G^{-1}+
\frac{1}{2!}v''(\bar{\varphi})G]+\frac{3}{4!}\hbar^2
v''''(\bar{\varphi})G^2\;\; ,
\label{g27}
\end{equation}
\noindent
which shows the classical contribution plus quantum corrections. In
Fig. 2 we present a contour plot of equipotential lines for the effective
potential. Its most interesting feature is the valley in the potential
energy surface which leads to a ``saddle point region'' (cf. below)
separating the two effective potential wells. Thus, our semiclassical
potential shares some similarity with the classical
``demonic potential'' studied in Ref. \cite{Heller2}. By analogy, we
expect also in our case energy dependent transitions between regular
and chaotic motion in the effective potential, which will be confirmed
in the next section.

Clearly, by opening the possibility to move into the semiclassical
$G$-direction, our system can roll around the hill at $\bar{\varphi} = G
= 0$, i.e. around the equivalent of the classical local maximum of the
potential at $\bar{\varphi} = 0$. This ``roll around'' represents the
quantum mechanical \underline{tunneling} through the given potential
barrier in the semiclassical TDHF approximation. Note, however, that
the limits $G\rightarrow 0$ and $\hbar\rightarrow 0$ do not commute
in eq. (\ref{g27}) due to the singular term $\sim \hbar/G$, which makes
the proper classical limit a nontrivial affair here as well as in
the equations of motion, eqs. (\ref{g21}) - (\ref{g24}).

Next, we observe that there are five fixed points of the flow
described by eqs. (\ref{g21}) - (\ref{g24}). They correspond to the three
extrema of the classical potential and are determined in the semiquantum
case by the conditions $\dot{\bar{\varphi}}=\dot{\bar{\pi}}=\dot{G}=
\dot{\sigma}=0$. These, by the equations of motion together with eq.
(\ref{g27}), become $\bar{\pi}=\sigma =\frac{\partial V}{\partial
\bar{\varphi}}=\frac{\partial V}{\partial G}=0$, in analogy to the
classical case.

A first solution corresponding to the center of the saddle point
region of the effective potential, cf. Fig. 2, is given by
$\bar{\varphi}=0$ and $G=f(\frac{2\mu^2}{3\lambda},\frac{1}{4 \lambda})$
with $$f(a,b)\equiv a+(b+a^3+[b^2+2ba^3]^{\frac{1}{2}})^{\frac{1}{3}}
+(b+a^3-[b^2+2ba^3]^{\frac{1}{2}})^{\frac{1}{3}}\;\; .  $$
For $\mu^2=1.0$, $\lambda = 0.06$ we find $G\approx 33.341$. However,
this turns out to be a small dip in the potential energy surface, since
there are two fixed points close by. They are given by
$$\bar{\varphi}=\pm(\frac{6\mu^2}{\lambda}-3G)^{\frac{1}{2}}\;\; ,
\;\;\;\mbox{with}\;\; G=\frac{2\mu^2}{3\lambda}(1+2\cos\frac{\alpha}{3})
\;\; ,\;\;\;\cos \alpha =1-(8\lambda (\frac{2\mu^2}{3\lambda})^3)^{-1}
\;\; .$$
For the above parameter values we find $\bar{\varphi}\approx\pm 0.106$,
$G\approx 33.330$. Finally, the fixed points corresponding to the
minima of the two deep effective potential wells are given by the
previous formulae for $\bar{\varphi}$ and $G$, however, replacing
$\alpha\rightarrow\alpha -\pi$ in the equation for $\cos\alpha$
only. Thus, we obtain here $\bar{\varphi}\approx\pm 9.947$, $G\approx
0.355$. Naturally, one may expect the trajectories to behave regularly,
if they are initiated with sufficiently small momenta $(\bar{\pi},\sigma)$
sufficiently close to the fixed points which correspond to (local)
minima of the effective potential.

Furthermore, note that the number of fixed points has increased by
two as compared to the classical limit. We conjecture that the number of
fixed points would keep increasing, if we would include corrections of
higher order in $\hbar$ into the effective action, eq. (\ref{g18}), and
therewith into the equations of motion. Equivalently, one would have to
go beyond the Gaussian ansatz for the wave function, eq. (\ref{g17}),
which will be further discussed in Section V.

Finally, we turn to the investigation of \underline{instabilities}
arising in the equations of motion, which might signal the onset of
chaotic behavior. It is well known that ordinary pertubation theory w.r.t.
the nonlinear coupling, i.e. $\lambda$ at present, would not give the
irregular break-up of KAM tori etc. for our conservative dynamical system
generated by the TDHF approximation \cite{Schuster,Ott}.
Therefore, we eliminate the momenta $(\bar{\pi},\sigma)$ from eqs.
(\ref{g21}) - (\ref{g24}) in order to rewrite them in second-order form
and, thus, to expose the analogy to coupled (potentially unstable)
oscillators. We obtain from eqs. (\ref{g21}), (\ref{g22}):
\begin{equation}
\ddot{\bar{\varphi}}+m^2_{eff}\bar{\varphi}=0\;\; ,
\label{g28}
\end{equation}
\noindent
with
\begin{equation}
m^2_{eff}\equiv\frac{1}{3!}\lambda\bar{\varphi}^2+\frac{1}{2}\lambda
G-\mu^2\;\; .
\label{g29}
\end{equation}
\noindent
{}From eqs. (\ref{g23}), (\ref{g24}) one finds:
\begin{equation}
\ddot{G}+M^2_{eff}G=\frac{1}{2}\frac{\dot{G}^2+1}{G}\;\; ,
\label{g30}
\end{equation}
\noindent
with
\begin{equation}
M^2_{eff}\equiv\lambda\bar{\varphi}^2+\lambda G- 2\mu^2\;\; .
\label{g31}
\end{equation}
\noindent
Obviously, both oscillators can become unstable due to exponentially
growing modes for a finite range of (initial) values for $\bar{\varphi}$
and $\sigma$. Both effective mass squared terms become negative for
\begin{equation}
0<G< \frac{2\mu^2}{\lambda}-\bar{\varphi}^2\;\; .
\label{g32}
\end{equation}
Note that the repulsive term on the r.h.s. of eq. (\ref{g30}) keeps
$G>0$ at all times. Thus, there is a region in the positive
$\bar{\varphi}$-$G$-halfplane, where we expect trajectories with
slightly different initial conditions (keeping for simplicity $\bar{\pi}=
\dot{\bar{\varphi}}=\sigma=\dot{G}=0$ initially) to diverge exponentially.

However, eq. (\ref{g32}) seems to provide at best a criterion for the
local sensitivity to initial conditions. In particular, for initial
conditions corresponding to high energy, i.e. sufficiently small
$G(t=0)$ according to eq. (\ref{g26}), one may expect the trajectories
to behave rather regularly, since the unstable region in the
$\bar{\varphi}$-$G$-plane by eq. (\ref{g32}) becomes a smaller and
smaller fraction of the accessible one with increasing energy. These
effects will be seen in our numerical results to be presented in the
following section.

\section{Numerical Results}

In this section we present the results of numerically studying the
solutions of the semiquantum equations of motion, eqs.
(\ref{g21}) - (\ref{g24}), for a variety of initial conditions.
However, we always fixed $\bar{\pi}=\sigma =0$ initially and chose
$G=0.1$ together with various values for $\bar{\varphi}$ to represent
initial configurations with varying total energy,
$E(\bar{\pi}=\sigma =0)=V(\bar{\varphi},G)$, cf. eq. (\ref{g27}).
Generally, we did not explore the full set of all initial
conditions at a given energy, which would require much more extensive
numerical calculations.

To begin with, we show in Fig. 3 the results for two energies which
are very close to the absolute minimum, $E_{min}\approx -24.2948$;
$E_{min}$ is obtained at the minima of the effective potential
$(\bar{\varphi}=\pm 9.947$, $G=0.355)$ discussed previously. Thus,
we observe in the left column of Fig. 3, which pertains to the lowest
energy, a rather simple regular motion characterized by the basic
oscillator frequency $\omega_0=\sqrt{2}$ and a second frequency at
twice this value. Note that the value of $\omega_0$ is obtained by
expanding the classical potential, eq. (\ref{g2}), around one of its
minima and identifying the coefficient of the harmonic term, i.e.
$\frac{1}{2}\omega^2_0=\mu^2=1.0$, at present. Together with the
frequency correlation function, cf. eqs. (\ref{g8}), (\ref{g12}),
and (\ref{g17}), we present two Poincar\'e sections in the phase space
planes of the two conjugate pairs of variables,
$\{\bar{\pi},\bar{\varphi}\}$ and $\{\sigma,G\}$, respectively.

Furthermore, a full trajectory is portrayed in the last row in Fig. 3.
Clearly, a slight increase of the energy, with the corresponding results
presented in the second column here, immediately leads to the
admixture of higher multiples of the basic frequency $\omega_0$.
In this case, the Poincar\'e sections still look fairly simple, but the
picture of a trajectory already shows a rather convoluted structure.
Obviously, the increasing complexity of trajectories can be expected
to be reflected in the more and more detailed fine structure of the
frequency correlation function as the energy is increased further.

Before we proceed, a few remarks are in order here concerning the
numerical solution of the equations of motion and the evaluation of the
frequency correlation function, in particular.

We have used {\it Mathematica}~\cite{MATH} to solve eqs. (\ref{g21}) -
(\ref{g24}) numerically. In particular, we employed the ND solve function
which iteratively and adaptively solves the equations to arbitrary accuracy
(up to the working precision of the computer, of course) which we chose to be
10 digits
in the final solution. The solutions for various initial conditions were
calculated for varying lengths of time depending on the complexity of the
trajectory. Thus, the results in Fig. 3 were obtained with 200 time units
($\hbar =c=1$)
while one of the Poincar\'e sections in Fig. 4 (see below) was obtained from
a solution of length 10,000 time units (the longest of our trajectories).

As mentioned earlier, one measure of the accuracy of our solution is
to compare the total energy $E$ at the beginning and ending of each
trajectory. For $E=-1.2661$ initially and the initial conditions stated
above, we obtain $E=-1.2660$ after 10,000 time units.

Evaluating the numerical solutions to eqs. (\ref{g21}) - (\ref{g24}) at
discrete time intervals,
we find the frequency correlation function by discrete Fourier transform,
i.e. by replacing the integral in eq. (\ref{g25}) with a sum
over the discrete time intervals of the trajectory. For our purposes
it was sufficient to take the discrete time interval $\Delta t=0.5$
which is more or less arbitrary. It is sufficient in the sense that the
discrete nature of $\rho (\varphi,\varphi;\omega)$ is clearly evident, or
not, depending on the particular value of the total energy and the initial
conditions. Tests with smaller values of $\Delta t$ on
selected trajectories produced the same low frequency spectrum.

The length of the trajectories, $\tau$, used to calculate
$\rho (\varphi,\varphi;\omega)$ was also somewhat arbitrary. $\tau$ must be
long enough to capture the basic low frequency structure. Again, our
objective was to observe the qualitative behavior of the frequency
correlation function rather than make a precise determination of its
content, and we found that trajectories of length
200 time units were sufficient for this purpose.
For the regular trajectories, increasing $\tau$ did not alter the
frequency spectrum, and for the trajectories that exhibited chaotic
character, increasing the trajectory length did not change the overall
character of the frequency correlation function.

Finally, we note that $\rho (\varphi,\varphi;\omega)$ as calculated above
is normalized to $N$ times its continuum value, eq. (\ref{g25}), where
$\tau=N \Delta t$.

To construct Poincar\'e sections, we simply evaluated each solution for a
particular total energy at an arbitrary fixed value of one of the four
variables while the others were allowed to vary. In practice, a small
variation in the fixed d.o.f. was allowed to reduce the number of
evaluations of the solution, i.e. our two-dimensional slice of phase
space had a small but finite thickness. Also, to observe the characteristic
phase space filling of the chaotic trajectories, $\tau$ was increased to as
much as 10,000 time units.

Next, we present in Fig. 4 characteristic Poincar\'e sections for
various higher energies. At $E=-12.44$ (initialized with $G=0.01$) we still
observe regular behavior, which is also obvious from the frequency
correlation function (not shown). Notice that the trajectory is still
completely confined to the right effective potential well, cf. Fig. 2,
where it was initialized. Clearly, one expects that different initial
conditions at the same energy would shift this Poincar\'e section w.r.t.
the coordinate axes and distort it as well. At $E=-6.61$ the structure
of the trajectories becomes much more complicated with the Poincar\'e
section showing indications of break-up of KAM tori. This is better
represented in the frequency correlation function, see. Fig. 5, where
there are still outstanding discrete lines. However, they appear already
considerably shifted w.r.t. to (integer multiples of) $\omega_0$.
Furthermore, there is now a broad low-level background at fractional
frequencies corresponding to the rather spiky character of the trajectory
itself. From the Poincar\'e section it is obvious that the trajectory
begins to run into the valley of the effective potential, cf. Fig. 2,
which is the semiclassical equivalent to (attempts at) quantum mechanical
tunneling to the other potential well $(\bar{\varphi}<0)$.

At $E=-1.266$ and $E=15.6$ the Poincar\'e sections in Fig. 4 both show
the ``stochastic sea outlining islands of stability'' which is the
typical feature of irregular motion resulting from the break-up of KAM
tori \cite{Schuster,Ott}. Note that the lower energy here still
corresponds to trajectories which classically could not get to the other
side of the barrier at $\bar{\varphi}=0$ (height $v(0)=0$), i.e. they have
to tunnel there. At $E=15.6$ trajectories run well above the classical
barrier and would yield regular motion in the classical limit. However,
both cases show semiquantum chaos as a result of quantum corrections.

Finally, in Fig. 5 the frequency correlation function is also shown
for two other energies, $E=-3.686$ and $E=201.8$, which reveal further
interesting features of our system. First, as expected, in the high
energy limit the behavior becomes regular again, since the effective
potential essentially becomes a single well with a perturbation of
decreasing influence at the center. This is obvious from the appearance
of the frequency correlation function. Presumably, its spectrum could be
related to the quantum mechanical energy levels of the anharmonic
$\bar{\varphi}^4$-oscillator relevant here at high excitation energy.

Secondly, and quite surprisingly indeed, we find at $E=-3.686$
regular behavior with a large number of discrete lines in the spectrum,
which are rather evenly spaced and of comparable strength. It seems
to originate from a resonance phenomenon leading to a fairly simple
\underline{closed trajectory} confined to one well, see Fig. 6. In
distinction to the cases of Fig. 3, where the motion results from the
quasi linear superposition of a few harmonic frequencies, at present
the trajectory appears simple despite being fully nonlinear. This
orderly behavior arises at an energy which lies in between cases with
much more irregular if not fully chaotic trajectories, cf. Fig. 5.
We did not yet investigate in full detail the sensitivity of this
phenomenon to changes of the energy control parameter etc. In particular,
it is an open question, whether other regular energy values like this
one exist within the irregular intermediate energy regime. Preliminary
investigations, however, show that this ``resonating trajectory'' is
even sensitive to where precisely it is initiated on an equipotential
contour, cf. Fig. 2. Presumably, we encounter a regime with stable KAM
tori side by side with irregular domains \cite{Schuster,Ott}, which
is already indicated by the Poincar\'e sections in Fig. 4.

The results presented here naturally raise a number of more general
questions. They will be discussed in the following section.

\section{Conclusions}

Presently, we have been studying semiquantum chaos in a simple
one-dimensional double-well oscillator model. In the Introduction we
explained the motivation for this and related studies by field theory
considerations and nonrelativistic many-body problems
\cite{Cooper,Jona-Lasinio,Elze,Heller}. In this context our present work
presents first steps towards a more detailed understanding of those more
complicated systems and their behavior in the semiclassical regime,
in particular.

Our work is based on the simple observation that a semiclassical
description (employing the Gaussian variational principle or the TDHF
approximation in our case, cf. Section III) necessarily involves the
introduction of additional degrees of freedom beyond the classical
ones, which may define the model. The corresponding additional
coordinate and conjugate momentum variables are related to quantum
fluctuations of the classical ones, i.e. they represent correlation
functions of varying order (cf. below) in the quantum mechanical
sense \cite{Pattanayak}. The classical and nonclassical degrees of
freedom are governed by a coupled set of first-order autonomous
differential equations, cf. eqs. (\ref{g21}) - (\ref{g24}). They are
necessarily highly nonlinear, which to a major extent is the case
independently of the particular dynamical model. Herein we find the
potential for deterministic semiquantum chaos in this effective
nondissipative Hamiltonian system.

Our numerical results presented in Section IV confirm the above
considerations, in general. In Section III we argued that the total
energy in our model, as determined, for example, by the parameters of
the initial state wave function, cf. eq. (\ref{g17}), serves as the
control parameter. Depending on the energy we find regular and chaotic
trajectories as solutions of the semiclassical equations of motion.
Besides Poincar\'e sections we mainly employed a suitably defined
frequency correlation function, cf. eqs. (\ref{g8}), (\ref{g25}),
which is related to the density matrix of the system to monitor its
behavior. Here the transition to chaos could be expected to show up
as a drowning of the discrete line structure deduced in Section II
in a rising broadband noise. This was clearly observed in the numerical
results, see, e.g., Fig. 5. However, quite surprisingly we found that
the behavior of the system as a function of the energy control parameter
changes in a rather complex way. Apparently, regular and irregular
regimes alternate with each other, where only the simplest cases
(low/high energy limit) can be easily understood. The simple-minded
stability analysis in Section III allows to outline qualitatively
some regime of irregular behavior, but is quantitatively not sufficient
to point out more of the detailed structure as a function of energy.

Clearly, our numerical studies show that there is room for more
surprises in the behavior of the system depending on the control parameter.
Furthermore, we did not fully explore the phase space of our system at a
given energy: Are all initial conditions equivalent in that the
corresponding trajectories are either regular or irregular? These
problems, being intimately related to the question of ergodicity in the
semiclassical model, can only be addressed through more extensive
numerical work, which is beyond the scope of our present paper.
Similarly, we did not investigate the nature of the transition to chaos
in this system, nor did we look into details of the control parameter
dependence of the frequency correlation function spectra.

However, our results together with the above considerations also
point towards several issues of more general interest, which we discuss in
the remaining part of this section.

It is convenient to introduce a formal representation of the exact
wave function and Schr\"odinger equation for any quantum mechanical system
which generalizes the Gaussian ansatz and variational equations of
motion, eqs. (\ref{g17}) and (\ref{g21}) - (\ref{g24}), repectively. We
rewrite the wave function,
\begin{equation}
\psi (x,t)=\exp f(x,t)\equiv\exp\sum\limits_{n=0}^{\infty}f_n(t)x^n\;\; .
\label{g33}
\end{equation}
\noindent
This ansatz, of course, can only be considered as formal, since we do
not know how to work out the normalization condition,
\begin{equation}
1=\int dx\;\psi^\ast\psi =\int dx\;\exp\{ 2\mbox{Re}f(x,t)\}\equiv
F[\mbox{Re}f_n (t)]\;\; ,
\label{g34}
\end{equation}
\noindent
in general. Nevertheless, pretending we know the highly nonlinear
functional $F$, we obtain a representation of the exact Schr\"odinger
equation, cf. eq. (\ref{g3}), as an infinite autonomous set of coupled
first-order differential equations,
\begin{eqnarray}
i\dot{f}_0        &=&-\frac{2\cdot 1}{2}f_2+\frac{i}{2}\frac{\dot{F}}{F}
\;\; , \label{g35} \\
i\dot{f}_1        &=&-\frac{3\cdot 2}{2}f_3+a\;\; , \label{g36} \\
i\dot{f}_2        &=&-\frac{4\cdot 3}{2}f_4+b\;\; , \label{g37} \\
i\dot{f}_3        &=&-\frac{5\cdot 4}{2}f_5+c\;\; , \label{g38} \\
i\dot{f}_4        &=&-\frac{6\cdot 5}{2}f_6+d\;\; , \label{g39} \\
i\dot{f}_{n>4}    &=&-\frac{(n+2)(n+1)}{2}f_{n+2}\;\; \label{g40} ,
\end{eqnarray}
\noindent
where we assumed for definiteness a simple polynomial potential,
$v(x)=ax+bx^2+cx^3+dx^4$. Several observations seem worth mentioning
here:
\begin{itemize}
\item[$\bullet$] The above exact representation of the full quantum
dynamics is not limited to one dimension and can be extended to field
theory using the functional Schr\"odinger picture, cf. \cite{Elze,Jackiw},
for example. However, the power series ansatz, eq. (\ref{g33}), is not
general enough or the most convenient one to cover all interesting
cases and is used here only for heuristic reasons.
\item[$\bullet$] There is an essential nonlinearity due to the
normalization condition which, however, appears only in the first of
the infinite set of coupled equations.
\item[$\bullet$] Different polynomial type interactions enter the
equations as different constant parameters only.
\end{itemize}

Based on these observations we conjecture the following. Any finite
truncation of sufficiently high order ($n\le N$) in such a nonlinear
expansion scheme for the full quantum dynamics, even of a classically
regular system, may lead to \underline{semiquantum chaos} (in the sense
presented in this paper) depending on the energy control parameter.
In general, we expect additional classical conservation laws to lead
to a higher dimensional control parameter space. Since the one and
only non-linearity originates from the wave function normalization
condition, cf. eqs. (\ref{g34}), (\ref{g35}), and enters the flow
equations in a completely different manner than any interactions, we
expect some still to be specified universality of the phenomena of
semiquantum chaos.

Finally, knowing that any exact wave function solution of the linear
Schr\"odinger equation cannot itself behave chaotically as discussed
in Section II, we are led to speculate that the infinite number of
linearly coupled correlation degrees of freedom represented by eqs.
(\ref{g35}) - (\ref{g40}) effectively restore the linear behavior of
the full quantum dynamics. The latter here was only undermined by a
single nonlinear constraint entering as a scale independent nonlinear
coupling into one of the flow equations!

This brings us to the final topic of our discussion.
In view of the linear character of the Schr\"odinger wave equation
and particularly for classically regular systems one might be tempted to
consider semiquantum chaos as a spurious effect caused by a truncation of
the full quantum dynamics as described above or any other bad
approximation thereof.

First of all, to illustrate the quality of the time-dependent Hartree-Fock
approximation we include here Fig. 7 showing some typical comparisons
between TDHF results and those of exact solutions of the Schr\"odinger
equation for given initial conditions. Overall the expectation value
of the position of the wave packet, for example, is always well described
for sufficiently short times. Here the scale should be set by the
characteristic time for the quantum mechanical spreading of the initial
wave packet. Clearly, this has to depend on the energy at present. Even
the long-time behavior is qualitively reproduced rather well in Fig. 7,
no matter whether the TDHF trajectories are regular or chaotic at a
particular energy. There is one notable exception here at $E=-1.266$,
where TDHF for the chosen initial conditions is fully tunneling by a
``roll-around'' as discussed in Section II, whereas the major part of
the exact wave packet stays is the initial well for long times.
Obviously, TDHF misses some of the interference structure of full
quantum mechanics. For further comparisons of TDHF and exact results in
different quantum mechanical models we refer to Ref. \cite{Cooper2}.

In any case, however, one may argue that for a \underline{closed
system} the full quantum description is always better than any
approximation. Thus, the way out of this apparent no-go situation, in
order to establish the physical relevance of semiquantum chaos, is to
allow for dissipative effects. Necessarily, one has to consider the
interaction of the quantum system under study with its
\underline{environment}. In this way, the observation of semiquantum
chaos is tied to the identification of suitable
\underline{open quantum systems} which through interactions with their
decohering environment \cite{Zurek} are driven into the semiclassical
regime.

Various steps in approaching a physically relevant dynamically
enforced classical limit (not naively setting $\hbar\approx 0$) have
been made \cite{Elze,Elze2,Zurek2,GH}, ranging in context from
cosmology through strong interactions, questions of foundations of
statistical mechanics, and the interpretation of quantum mechanics.
Presently, with the examples discussed in the Introduction in mind,
it seems most relevant to find systems, e.g. in solid state physics
(``quantum dots'' \cite{Reed}), where environment-induced
quantum decoherence and consequently semiquantum chaos become
experimentally accessible.

Based on the example treated in Ref. \cite{Elze}, a quantum particle
showing essentially classical behavior due to its interaction with a
suitable quantized radiation field, the following model suggests itself
for further study: the double-well oscillator governing a quantum
particle coupled to a decohering radiation field. Before any detailed
results, it would be interesting to see the effect of a deliberately
chosen environment on the relevant set of flow equations, such as
eqs. (\ref{g35}) - (\ref{g40}) above. We plan to address some of these
fascinating problems in a future publication.
\vskip 0.2cm
\noindent
{\bf Acknowledgements:} This work was supported in part by
the University of Arizona (Tucson).

\newpage
\underline{\bf Figure Captions:}
\vspace{1.5cm}
\begin{itemize}
\item[Fig. 1:] The double-well oscillator potential, eq. (\ref{g2}),
with $\mu^2 = 1.0, \lambda = 0.06$. Also shown is a typical classical
phase space trajectory ($\phi\equiv\bar{\varphi}$).
\item[Fig. 2:] The effective semiquantum potential, eq. (\ref{g27}),
for $\phi\equiv\bar{\varphi}\ge 0$ (V is symmetric in $\phi$). The inset
shows the detailed shape of the contours close to the positive minimum
(i.e. fixed point) of V.
\item[Fig. 3:] For two low energies (left/right column) the following
TDHF results are shown (top to bottom): frequency correlation function,
Poincar\'e sections in the two different coordinate/conjugate momentum
planes, full trajectory (see text for details).
\item[Fig. 4:] Poincar\'e sections at various values of the energy
(control parameter) illustrating the transition from regular to
chaotic motion.
\item[Fig. 5:] Frequency correlation functions at various energies.
Semiquantum chaos shows up as broadband noise in the expected line
spectrum. There is an unexpected resonance effect at $E=-3.68601$.
\item[Fig. 6:] The nonlinear resonance trajectory discussed in the
text.
\item[Fig. 7:] Comparisons between exact solutions of the time-dependent
Schr\"odinger equation (full lines) and TDHF results
(symbols) at various energies. Shown here is the expectation value of
the position of the particle, which is initially a Gaussian wave packet.
\end{itemize}

\newpage
\begin{thebibliography}{99}
\bibitem{Schuster} H. G. Schuster, ``Deterministic chaos: an
introduction'', 2nd ed. (VCH, New York/Weinheim 1989).
\bibitem{Cooper} F. Cooper, J. F. Dawson, D. Meredith and H.
Shepard, Phys. Rev. Lett. \underline{72} (1994) 1337.
\bibitem{Pattanayak} A. K. Pattanayak and W. C. Schieve, Phys.
Rev. Lett. \underline{72} (1994) 2855.
\bibitem{Jona-Lasinio} G. Jona-Lasinio and C. Presilla, Phys.
Rev. Lett. \underline{68} (1992) 2269.
\bibitem{Ryder} L. H. Ryder, ``Quantum Field Theory'' (Cambridge
University Press, Cambridge 1991).
\bibitem{Elze} H.-Th. Elze, Nucl. Phys. \underline{B436} (1995) 213;
Nucl. Phys. B (Proc. Suppl.) \underline{39B,C} (1995) 169; CERN-TH.
7297/94, hep-th/9406085.
\bibitem{Jackiw} R. Jackiw and A. Kerman, Phys. Lett. \underline{71A}
(1979) 158.
\bibitem{Heller} E. J. Heller, J. Chem. Phys. \underline{62} (1975)
1544.
\bibitem{Heller2} E. J. Heller and R. L. Sundberg, in ``Chaotic
Behavior in Quantum Systems'', edited by G. Casati (Plenum Press, New
York 1985), p. 255.
\bibitem{Carruthers} P.A. Carruthers and F. Zachariasen, Rev.
Mod. Phys. \underline{55} (1983) 245.
\bibitem{Ott} E. Ott, ``Chaos in dynamical systems'' (Cambridge
University Press, Cambridge 1993).
\bibitem{Elze2} H.-Th. Elze, CERN-TH. 7372/94, hep-ph/9407377; in
``Quantum Infrared Physics''; edited by H. M. Fried and B. M\"uller (World
Scientific, Singapore 1995), p. 95.
\bibitem{Grabert} H. Grabert, P. Schramm and G.-L. Ingold, Phys. Rep.
\underline{168} (1988) 115.
\bibitem{Cooper2} F. Cooper, S.-Y. Pi and P. N. Stancioff,
Phys. Rev. \underline{D34} (1986) 3831.
\bibitem{MATH} S. Wolfram, ``Mathematica'' (Addison Wesley, Reading MA 1991).
\bibitem{Zurek} W. H. Zurek, Phys. Today \underline{44}, No. 10
(1991) 36; \\
R. Omn\`es, Rev. Mod. Phys. \underline{64} (1992) 339; \\
H. D. Zeh, Phys. Lett. \underline{A172} (1993) 189.
\bibitem{Zurek2} W. H. Zurek and J. P. Paz, Phys. Rev. Lett.
\underline{72} (1994) 2508.
\bibitem{GH} M. Gell-Mann and J. B. Hartle, Phys. Rev. \underline{D47}
(1993) 3345.
\bibitem{Reed} M. A. Reed, Sci. American, Jan. 1993 (1993) 98.
\end{thebibliography}
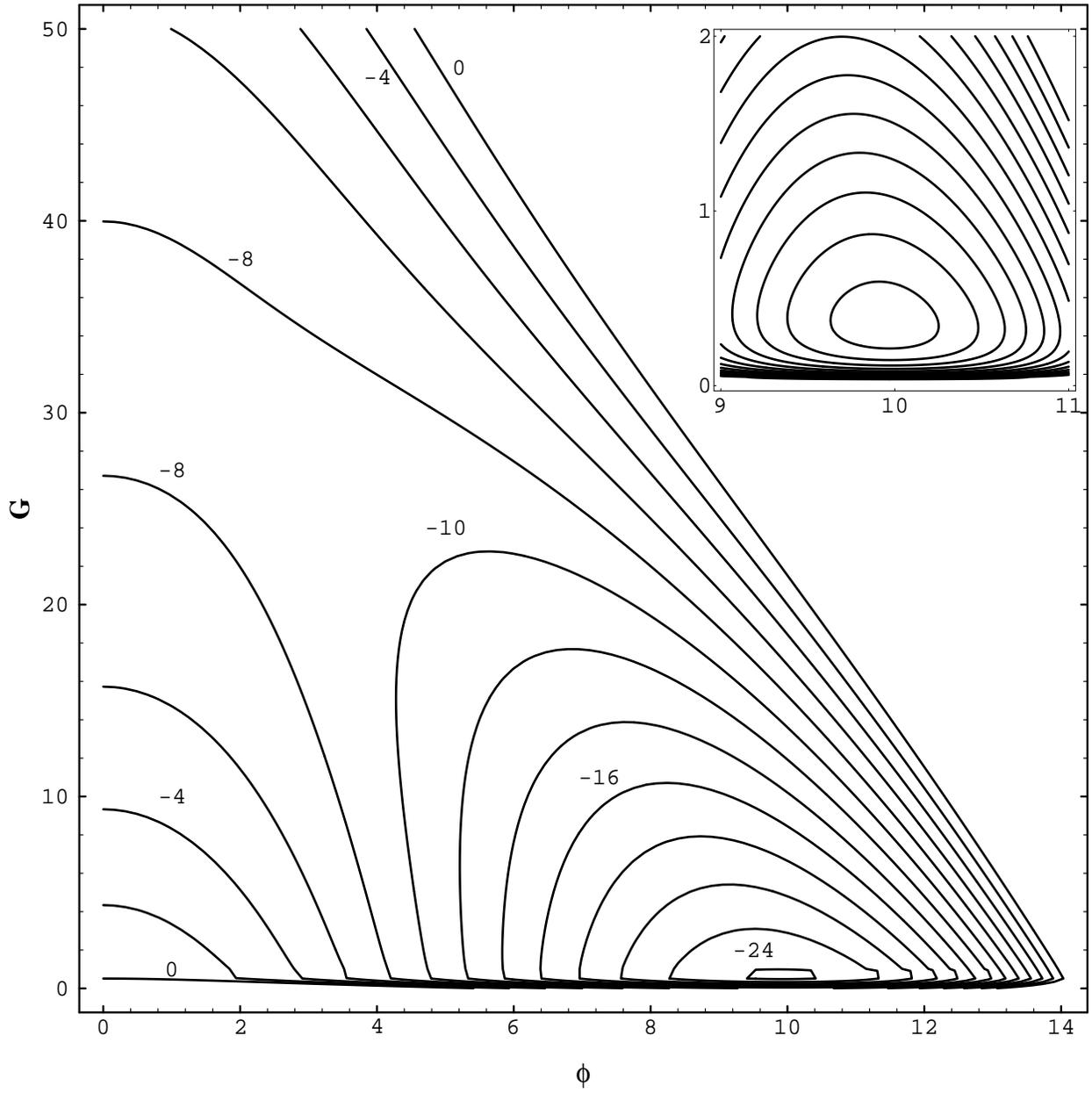
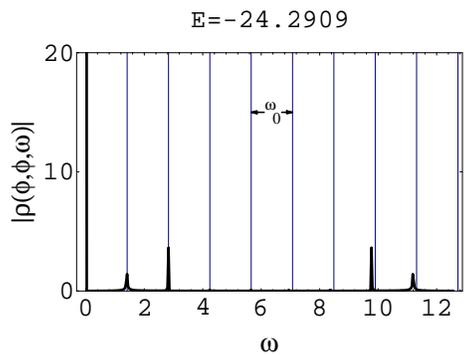
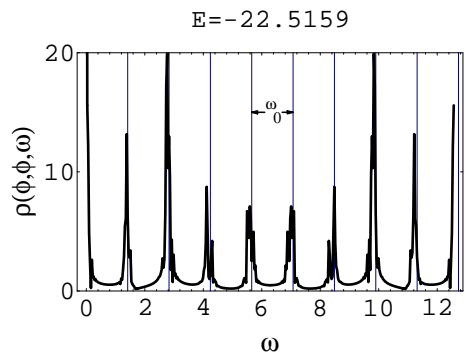
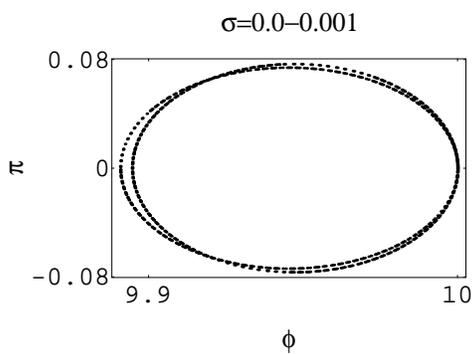
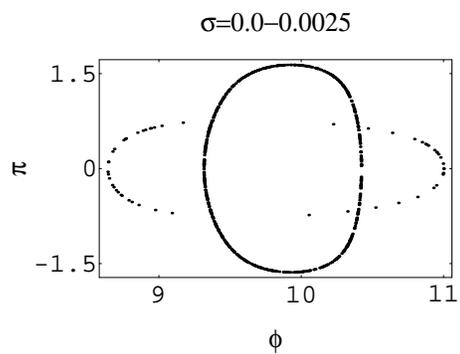
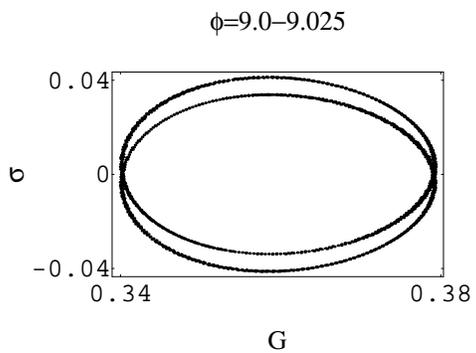
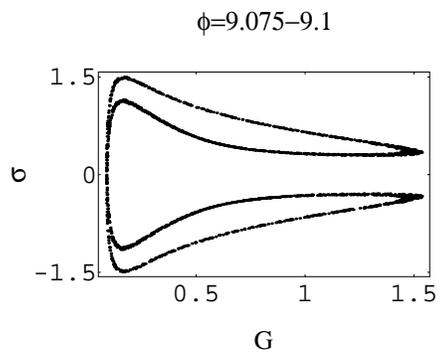
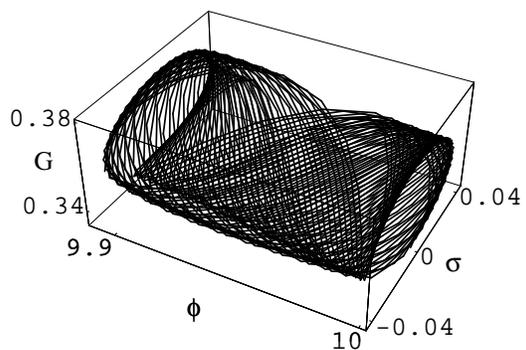
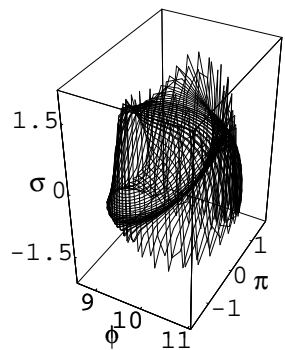
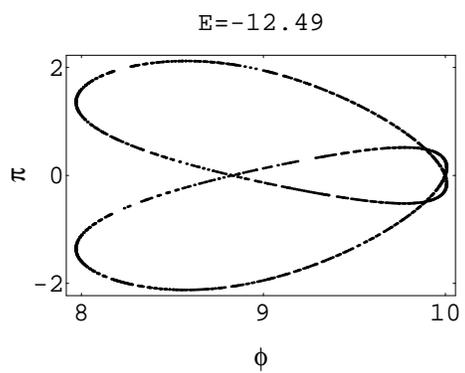
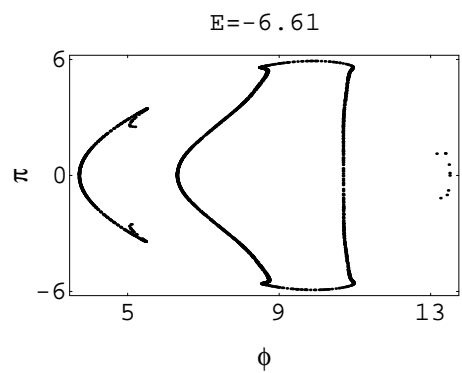
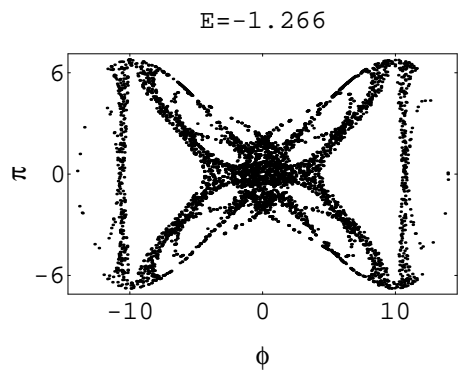
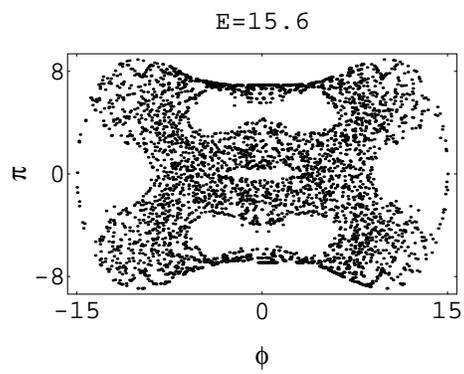
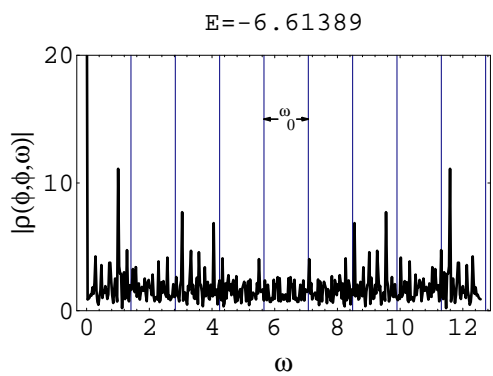
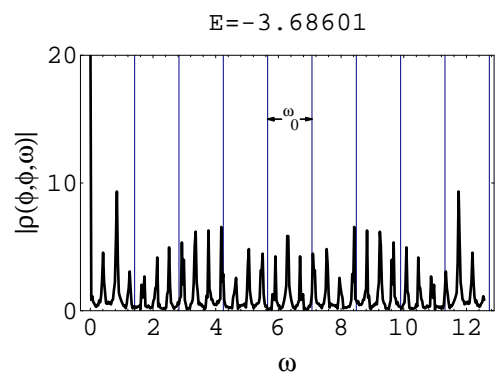
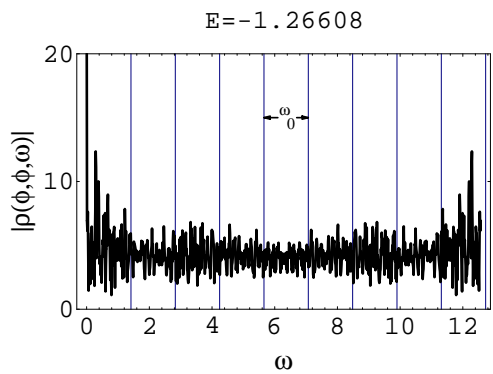
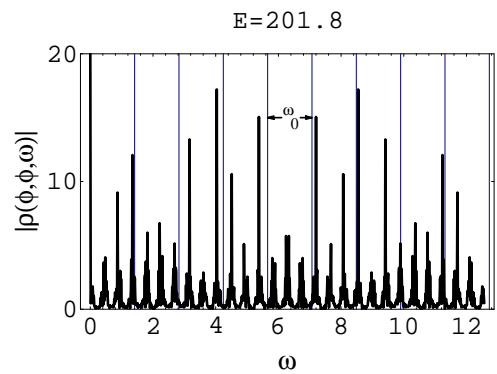
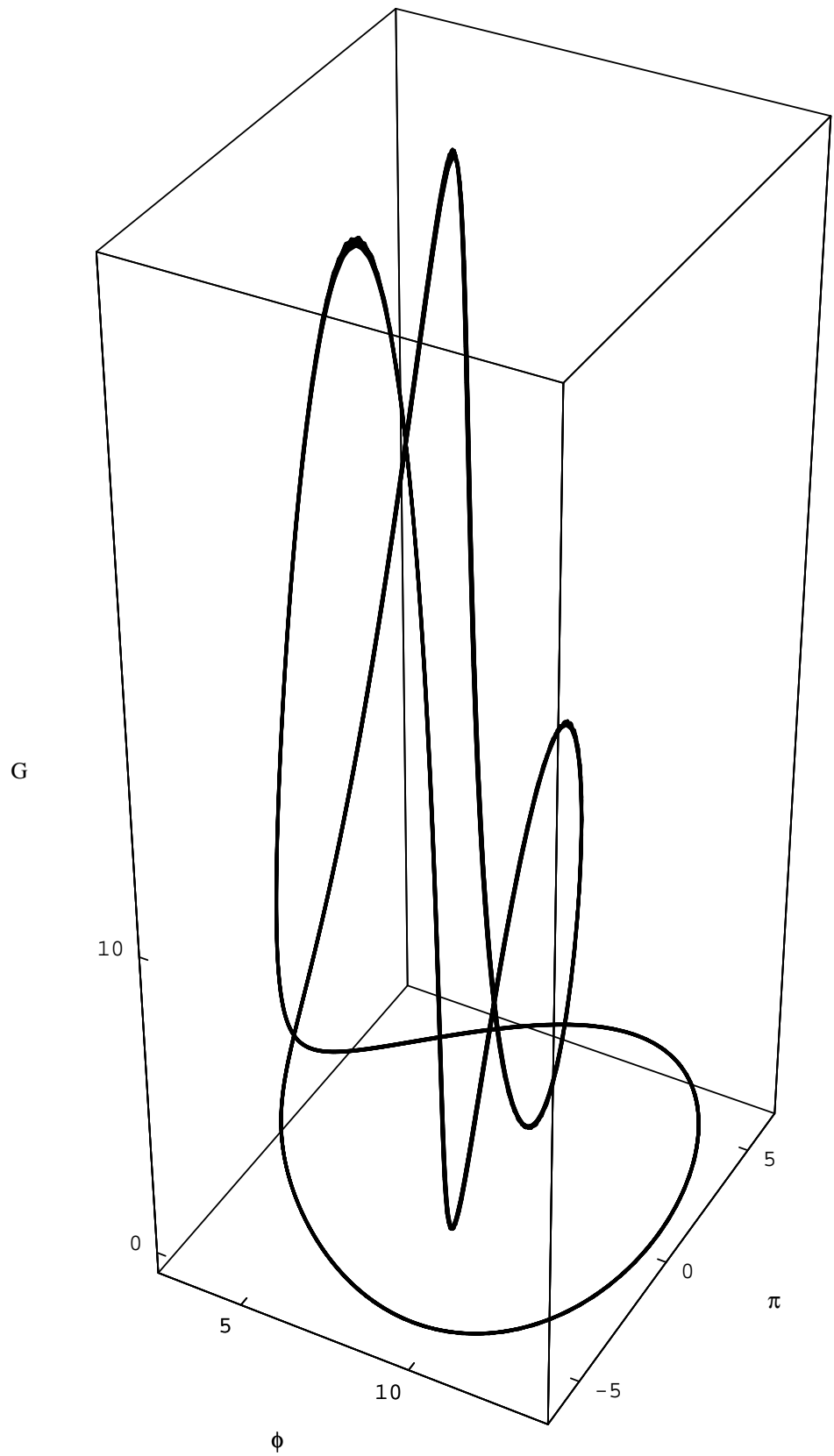
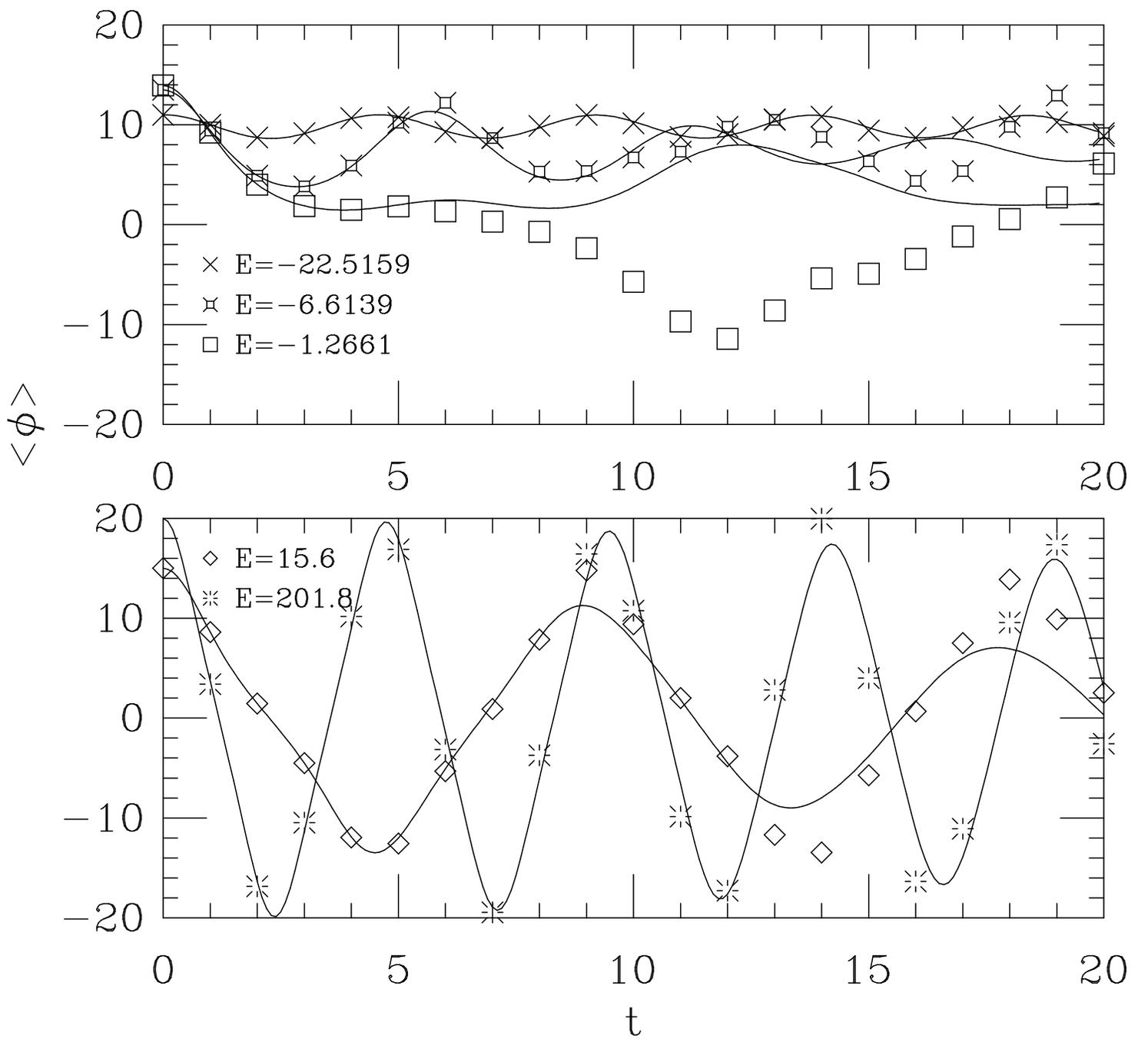
\end{document}